\documentclass{jfm}
\usepackage{graphicx}
\usepackage{epstopdf, epsfig}
\usepackage{amsmath}
\usepackage{float}
\usepackage[dvipsnames]{xcolor}
\usepackage{natbib}
\usepackage{hyperref} 
\usepackage{bm}
\usepackage{orcidlink}

\renewcommand{\d}		{\partial}

\newcommand{\dd}	    {{\rm d}}

\shorttitle{Momentum, energy, and vorticity balances in water waves}
\shortauthor{A. Blaser, R. Benamran, A. B. Villas Bôas, L. Lenain, N. Pizzo}

\title{Momentum, energy and vorticity balances in deep-water surface gravity waves}

\author{Aidan Blaser\aff{1}, Rapha{\"e}l Benamran\aff{2}, Ana B Villas Bôas\aff{3}, Luc Lenain\aff{1} and Nick Pizzo\aff{4}\corresp{\email{nicholas.pizzo@uri.edu}}}

\affiliation{\aff{1}Scripps Institution of Oceanography, University of California San Diego,
La Jolla, CA 92037, USA \aff{2}Department of Physics, Brown University, Providence, RI 02912 , USA \aff{3}Department of Geophysics, Colorado School of Mines, Golden, CO 80401, USA\aff{4}Graduate School of Oceanography, University of Rhode Island, Narragansett, RI, USA 02881} 
\begin{document}

\maketitle

\begin{abstract}



The particle trajectories in irrotational, incompressible and inviscid deep-water surface gravity waves are open, leading to a net drift in the direction of wave propagation commonly referred to as the Stokes Drift, which is responsible for catalysing surface wave-induced mixing in the ocean and transporting marine debris. A balance between phase-averaged momentum density, kinetic energy density and vorticity for irrotational, monochromatic and {\color{black} spatially periodic} two-dimensional water waves is derived by working directly within the Lagrangian reference frame, which tracks particle trajectories as a function of their labels and time. This balance should be expected as all three of these quantities are conserved following particles in this system. Vorticity in particular is always conserved along particles in two-dimensional inviscid flow, and as such even in its absence it is the value of the vorticity which fundamentally sets the drift, which in the Lagrangian frame is identified as the phase-averaged momentum density of the system. A relationship between the drift and the geometric mean water level of particles is found at the surface {\color{black} which highlights connections between geometry and dynamics.} 
Finally, an example of an initially quiescent fluid driven by a wavelike pressure disturbance is considered, showing how the net momentum and energy from the surface pressure disturbance transfer to the wave field, recognizing the source of the mean Lagrangian drift as the net momentum required to generate an irrotational surface wave by any conservative force.

\end{abstract}


\section{Introduction}

Deep water surface gravity waves are ubiquitous in the global oceans, and affect the transport of heat, momentum and mass both along and across the air-sea interface \citep{vanSebille2020,Melville1996,Deike2022}.
One crucial property of irrotational deep water waves is that the particle trajectories are not closed leading to a net drift in the direction of wave propagation commonly referred to as the Stokes Drift \citep{Stokes1847}.
Formally, the Stokes drift is defined as the difference between the mean Lagrangian and mean Eulerian currents,
\begin{equation}
    \bm{U}_{S} \equiv \overline{\bm{u}_L} - \overline{\bm{u}_E} \, ,
\end{equation}
where $\bm{u}_L$ and $\bm{u}_E$ represent the Lagrangian and Eulerian currents respectively, and the overline indicates a time mean in each reference frame over a wave period. One often neglected issue with this definition is the validity of taking the difference between two quantities in \emph{different reference frames} with \emph{different dependent variables} and more importantly \emph{different definitions of averaging} as the Lagrangian and Eulerian periods are not equal \citep{LH1986}. To avoid this confusion, we will instead use the term `wave-induced mean Lagrangian drift' to refer to the mean \emph{Lagrangian} velocity of fluid particles over the Lagrangian wave period.

The wave-induced mean Lagrangian drift modulates upper ocean currents, affects the transport of buoyant pollutants, plankton and marine debris {\color{black}\citep{dibenedetto_ouellette_koseff_2018}} and enhances vertical mixing via Langmuir circulation \citep{Craik1976, Belcher2012, Wagner2023}.
There is also evidence that this drift, or mean Lagrangian momentum density, can help with the interpretation of many central geometric, kinematic and dynamic properties of surface waves \citep{Pizzo2023}. Despite the elapse of over 175 years since its discovery, there is still confusion regarding the origins and interpretation of the wave-induced mean Lagrangian flow for irrotational surface gravity waves.
Most derivations, including that of \cite{Stokes1847}, calculate the magnitude and direction of the drift from an asymptotic integration of the kinematic condition relating the Eulerian and Lagrangian velocities, which simply states that at a fixed point in time and space, the Eulerian and Lagrangian velocities are equal
\begin{equation}
    \frac{\dd \bm{x}(t)}{\dd t} = \bm{u}_{E}(\bm{x}(t),t) \, ,
\end{equation}
since at a fixed time a particle's location is coincident with a fixed point in space. Thus the particle trajectories within a wave, which are fundamentally Lagrangian quantities, are typically derived from the Eulerian velocity fields.
When done in this way, the mean Lagrangian drift appears to simply fall out of the math, and physical explanations for its existence tend to come \emph{post-factum}.
Why, fundamentally, should progressive irrotational surface waves induce a mean motion of water?
What sets its magnitude and direction? Lastly, how is the mean Lagrangian drift related to other quantities such as vorticity and energy density? Such questions are the primary aim of this paper.

In section 2, we introduce the governing equations and relevant conditions for solving irrotational, incompressible, {\color{black}spatially periodic} and inviscid two-dimensional {\color{black} progressive} deep-water surface gravity waves in the Lagrangian reference frame, which tracks the trajectories of individual fluid parcels as a function of labelling coordinates. In this frame, the wave-induced mean Lagrangian drift is explicitly written as the average velocity of fluid parcels, and is thus identified as the mean Lagrangian momentum density of the system, physically motivated when one recalls momentum being equivalent to mass flux.
In section 3, through an investigation of these equations, we show how the mean momentum density, or equivalently the drift, is related to the vorticity and energy density of irrotational water waves. Despite the flow being completely irrotational, it is precisely this {\color{black} very} strict condition {\color{black} of irrotationality} that dynamically mandates a sheared mean Lagrangian drift, and thus we emphasize that even in irrotational flow it is the vorticity that sets the magnitude and direction of the drift. We then dynamically connect the mean momentum and kinetic energy densities, showing that all monochromatic, irrotational and progressive waves with nonzero kinetic energy require a nonzero mean Lagrangian drift. Finally, we highlight a connection between the mean potential and mean kinetic energy densities through the mean pressure using a Bernoulli equation in the Lagrangian frame first outlined by \cite{Pizzo2023}.

To further explore the dynamic relationship between these variables, in section 4 we consider the momentum and energy budgets within the physically motivated Lagrangian reference frame. To show how the wave-induced mean Lagrangian drift emerges from an initially quiescent flow, we consider a simple example where a still surface is resonantly driven by a wavelike pressure forcing.
Through the integral momentum budgets, we find that all the momentum transferred to the waves from the surface forcing goes into the mean momentum, or mean Lagrangian drift. The same is shown to be true for the total energy. {\color{black} Finally, while the main body of this work concerns irrotational flow due to its prevalence in the literature, one can turn to the appendix for a generalization of each major result for rotational flow.}

\section{The Lagrangian description of water waves}
Lagrangian quantities track evolution following fixed fluid particles. Thus a complete two-dimensional Lagrangian description of a fluid {\color{black} domain} requires calculating particle trajectories $\bm{x}(\alpha,\beta,\tau)$ {\color{black} for each fluid element} as a function of {\color{black} continuous} particle labelling coordinates $(\alpha,\beta)$ and time $\tau$. Note that we distinguish $\tau$ from $t$ to emphasize that the partial derivative with respect to $\tau$ holds particle labels fixed. We will equivalently indicate such derivatives with an overhead dot. The particle trajectories $\bm{x}(\alpha,\beta,\tau)$ represent a general time-dependent mapping between label space $(\alpha,\beta)$ and physical space $(x,y)$ {\color{black}which must be invertible because no two particles can occupy the same physical location at the same time: that is, the function}

\begin{equation}
    \mathcal{J} \equiv \frac{\partial (x,y)}{\partial(\alpha,\beta)} = x_\alpha y_\beta - x_\beta y_\alpha \, ,
\end{equation}
{\color{black} 
must not be equal to zero anywhere in the domain. Here subscripts indicate partial derivatives.}
{\color{black}The determinant of the Jacobian matrix of this mapping, $\mathcal{J}$, also allows us to easily change variables of differentiation.} 
For example, the two dimensional incompressiblity condition in the Eulerian frame is denoted as
\begin{equation}
    u_x + v_y = 0 \, ,
\end{equation}
which can be mapped to the Lagrangian frame through the following series of steps
\begin{multline}
    0 = u_x + v_y = \frac{\d (u,y)}{\d (x,y)} + \frac{\d (x,v)}{\d (x,y)} = \frac{1}{\mathcal{J}}\bigg( \frac{\d (u,y)}{\d (\alpha,\beta)} + \frac{\d (x,v)}{\d (\alpha,\beta)}\bigg) =\\
    \frac{1}{\mathcal{J}}\bigg( \frac{\d (\dot{x},y)}{\d (\alpha,\beta)} + \frac{\d (x,\dot{y})}{\d (\alpha,\beta)}\bigg) = \frac{1}{\mathcal{J}}\frac{\d}{\d \tau} \frac{\d (x,y)}{\d (\alpha,\beta)} = \frac{1}{\mathcal{J}}\dot{\mathcal{J}} = 0 \, .\label{continuity}
\end{multline}
\noindent
Therefore incompressible flow requires that the $\mathcal{J}$ be time independent. 
We could have determined this condition without calculation by remembering that the $\mathcal{J}$ determines how infinitesimal areas are mapped between label space and physical space.
A small collection of particles $\dd \alpha \, \dd \beta$ must enclose the same physical area $\mathcal{J}^{-1}\dd x \, \dd y$ for all time or else the flow would be allowed to compress.

Since we are considering inviscid flow, the Euler equations will suffice for our treatment.
Upon conversion to the Lagrangian frame they become \citep[Art. 15]{LAMB1932}
\begin{align}
        \mathcal{J} \ddot{x} + p_\alpha y_\beta - p_\beta y_\alpha = 0 \, , \label{"xEuler"}\\ 
        \mathcal{J} \ddot{y} + p_\beta x_\alpha - p_\alpha x_\beta + \mathcal{J}g = 0 \label{"yEuler"} \, ,
\end{align}
where $p$ represents the pressure, and $g$ the acceleration due to gravity. 
Note that in the Lagrangian frame the {\color{black}nonlinear terms} arise in the pressure terms and not in the inertia terms, in contrast to the Eulerian reference frame. 

While incompressibility provides a constraint on our mapping between label space and physical space, there is still tremendous freedom in how we label our particles; this is known as the particle relabelling symmetry, and represents a gauge freedom of fluid mechanics. {\color{black}The conserved quantity associated with this gauge freedom is the vorticity \citep{Salmon1988}}. Just as in electromagnetism, this gauge can be conveniently chosen to {\color{black} simplify} computations, but we leave it general for now.

Here, our physical system amounts to solving equations \eqref{"xEuler"}, \eqref{"yEuler"} for variables $(x,y,p)$ as functions of $(\alpha,\beta,\tau)$, subject to the incompressibility condition \eqref{continuity} for a given labelling gauge choice.
To close the system, we impose boundary conditions at the free surface and the bottom.
As part of our labelling freedom, we label particles at the surface with $\beta = 0$, which makes the evaluation of surface quantities straightforward. This is equivalent to saying that our domain in label space is just the lower half plane, which is much simpler to work with both theoretically and numerically. This is in contrast to in the Eulerian frame where the domain is bounded above by the free surface $\eta(x,t)$, which is itself a dependent variable of the system and not known \emph{a priori}.
Thus our surface boundary condition, equivalent to the dynamic boundary condition in Eulerian coordinates, simply states that pressure must vanish at the surface, i.e.
\begin{equation}
    p(\beta = 0) = 0 \, ,
\end{equation}
which is just another way of saying that the wave is unforced.
We examine what happens when this condition is relaxed in a later section.
The bottom boundary condition states that the vertical velocity must vanish as we tend towards the infinitely deep impermeable bottom
\begin{equation}
    \dot{y}(\beta = -\infty) = 0 \, .
\end{equation}

As a final point, the fact that the domain in label space is time independent also means that all points initially within the domain remain there. This is in contrast to the Eulerian frame, where certain points, such as those with $y=0$, are outside of the fluid part of the time, and therefore taking temporal averages at these points becomes ill-defined. The implications of these Eulerian averages are discussed further below.


\section{Drift in relation to vorticity, momentum and energy}

Up to this point we have neglected to mention the vorticity of these waves.
While it has been long known that there exists an exact solution to the above system in which particles undergo purely circular trajectories, these \cite{Gerstner1802} waves have a nonvanishing vorticity \citep[Art. 251]{LAMB1932}. 
When surface waves are generated from a state of rest by a (conservative) pressure gradient force, they are irrotational \citep{Phillips1977}.
We can compute the vorticity in Lagrangian coordinates by a simple mapping between reference frames
\begin{equation} \label{"vorticity"}
    q \equiv v_x - u_y =  \frac{\d (\dot{x},x)}{\d (x,y)} + \frac{\d (\dot{y},y)}{\d (x,y)} = \frac{1}{\mathcal{J}} \bigg(\frac{\d (\dot{x},x)}{\d (\alpha,\beta)}+ \frac{\d (\dot{y},y)}{\d (\alpha,\beta)} \bigg) \, ,
\end{equation}
so that irrotational flow requires
\begin{equation}\label{"Vorticity"}
    q\mathcal{J}  = \dot{x}_\alpha x_\beta - \dot{x}_\beta x_\alpha + \dot{y}_\alpha y_\beta - \dot{y}_\beta y_\alpha = 0 \, .
\end{equation}
Recall that in two-dimensional inviscid flow, vorticity is materially conserved along particles, e.g.
\begin{equation}
    \dot{q} =  0 \, .
\end{equation}
{\color{black} However, this does not extend to three dimensions, where vorticity is no longer materially conserved on particles (due to vortex tilting and stretching), but is instead conserved on one-dimensional vortex lines. In two-dimensions, these lines collapse to a point because they are assumed to extend indefinitely in a direction orthogonal to the plane. {\color{black}Here, we restrict ourselves to two dimensional wave fields with no variation in the transverse direction. } Applications to finite extent wave packets in both two and three dimensions is under current investigation by the authors  \citep[see also][]{Pizzo2021}.}

\subsection{Series expansions}

We consider {\color{black} permanent, progressive, spatially periodic and} monochromatic waves {\color{black}in two dimensions} and expand our trajectories in a series {\color{black} following \cite{Clamond2007,Pizzo2023}} as
\begin{equation}\label{expansions}
    x = \alpha + U(\beta)\tau + \sum_{n=1}^\infty x_n(\beta)\sin(\theta_n) \, , \qquad y = \beta + y_0(\beta) + \sum_{n=1}^\infty y_n(\beta) \cos(\theta_n) \, ,
\end{equation}
with $\theta_n = nk({\color{black}\alpha} - (c-U(\beta))\tau)$, $k$ the wavenumber, $c$ the phase speed, $U(\beta)$ the explicit mean Lagrangian drift, and $y_0(\beta)$ the mean water level, a parameter explored further in section 3.3. {\color{black}The justification for these expansions is as follows. First, we expand about a rest state $(x=\alpha,y =\beta)$. We then include an explicit steady mean Lagrangian drift in $x$ and a mean water level in $y$, which will be constrained by the equations of motion and boundary conditions. Due to the horizontal periodicity of the wave profiles, a Fourier series expansion is best suited to represent the wavelike orbital motion of the fluid particles. Because the waves are permanent in shape, the Fourier coefficients can depend  only on the vertical label $\beta$. }Note that the intrinsic frequency, Doppler-shifted by $U(\beta)$, is needed to remove secular terms at higher orders (see \citealp{Clamond2007} {\color{black}discussing} \citealp{buldakov2006new}).
Inserting these expansions into the irrotational condition \eqref{"Vorticity"} and taking the time averaged component yields
\begin{equation}
    \color{black}-U'(\beta) + (c-U(\beta))\sum_{n=1}^\infty n^2 k^2 (x_n'(\beta)x_n(\beta) + y_n'(\beta)y_n(\beta)) = 0 \, ,
\end{equation}
{\color{black}which can be integrated as}
\begin{equation} \label{"drift"}
    U(\beta) = \dfrac{\displaystyle \frac{c}{2} \sum  n^2 k^2(x_{n}^2 + y_{n}^2)}{\displaystyle 1 + \frac{1}{2} \sum n^2 k^2 (x_{n}^2 + y_{n}^2)} \, ,
\end{equation}
a result first found by \cite{Pizzo2023} which shows what form the drift must take to maintain irrotational flow. {\color{black} The integration constant is zero since we are considering the frame where the fluid velocity vanishes at depth.}
While this relation just comes from a dynamic constraint, it shows that, given an expansion of the form \eqref{expansions}, and so long as a wave is present (e.g. $x_n$,$y_n \neq 0$ for some $n$) there must be a positive definite mean Lagrangian drift $U(\beta)$ for the flow to stay irrotational.
While this line of reasoning explains why a sheared mean flow is needed if an irrotational wave is present, we still lack a physical mechanism for its origin.
To that end, we next turn to an investigation of wave dynamics.




\subsection{Drift and Energy}

Kelvin's circulation theorem states that the circulation of a material contour
\begin{equation}
    \Gamma \equiv \oint \bm{u} \cdot \dd \bm{\ell} \, ,
\end{equation}
is conserved following the flow (i.e. $\dot{\Gamma}=0$). A lesser known Lagrangian version of this theorem \citep[see][eq. 4.12]{Salmon1988} equivalently defines the circulation as
\begin{equation}\label{gammaDef}
    \Gamma = \oint \bm{A} \cdot \dd \bm{\alpha} \, , \qquad \bm{A} \equiv \dot{x} \bm{\nabla_{\bm{\alpha}}} x + \dot{y} \bm{\nabla_{\bm{\alpha}}}y \, , 
\end{equation}
in two dimesions where $\bm{\nabla_{\bm{\alpha}}} = (\partial_\alpha,\partial_\beta)$ is the gradient operator in label space. From this it is clear that
\begin{equation}\label{equivalence}
    \bm{u} \bm{\cdot} \dd \bm{\ell} = \bm{A} \bm{\cdot} \dd \bm{\alpha} \, ,
\end{equation}
which proves how these are equivalent representations of the circulation. For irrotational flow, we can always write the Eulerian velocity $\bm{u}$ as the gradient of a scalar velocity potential $\phi$. By the chain rule, we can show
\begin{equation}
    \bm{\nabla} \phi \bm{\cdot} \dd \bm{\ell} = \bm{\nabla_{\alpha}} \phi \bm{\cdot} \dd \bm{\alpha} \, ,
\end{equation}
which, when compared with \eqref{equivalence} shows that for irrotational flow $\bm{A}$ is just the gradient of the velocity potential $\phi$ in label space.

\begin{figure}
    \centering
    \includegraphics[width=8cm]{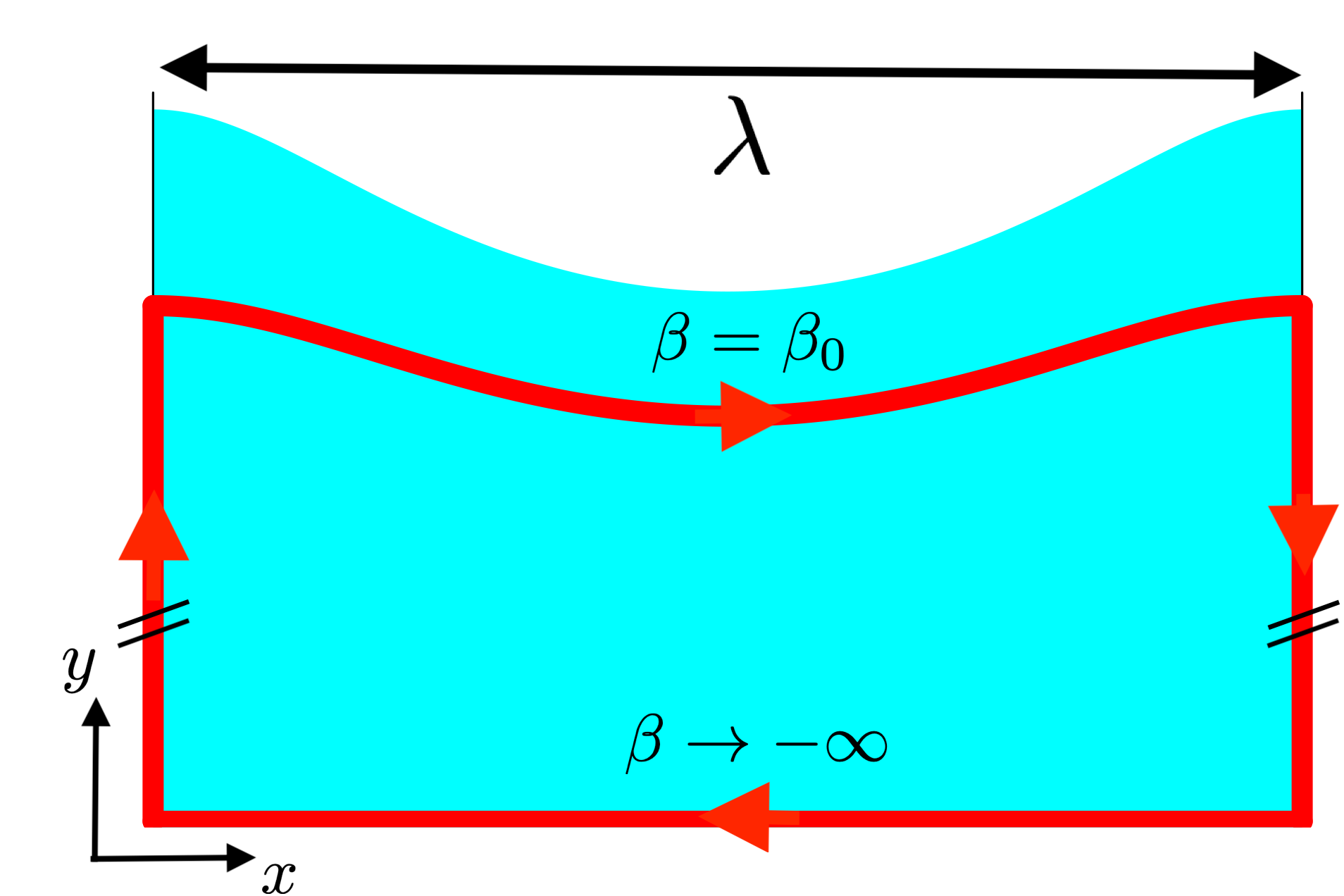}
    \caption{Schematic of a potential closed material loop for periodic, progressive waves (red). The top contour is a material line of constant vertical label $\beta = \beta_0$. Because $\vec{A}$ is $\lambda$-periodic, which can be shown from \eqref{expansions}, the side contours cancel. Due to our infinite bottom condition, $\bm{A}$ vanishes as we approach the bottom and there are no contributions there. Note the clockwise orientation used.}
    \label{fig:Kelvin}
\end{figure}

What makes the Lagrangian representation particularly interesting here is that the material loop in label space is fixed in time by definition, so Kelvin's circulation theorem reduces to
\begin{equation}\label{kelvin}
    \frac{\d \Gamma}{\d \tau} = \oint \frac{\partial \bm{A}}{\partial \tau} \bm{\cdot} \dd \bm{\alpha} = 0 \, ,
\end{equation}
for any closed loop in a potentially rotational fluid. However, if we constrain ourselves to irrotational flows, we have
\begin{equation}\label{GammaLag}
    \Gamma = \oint \bm{A} \bm{\cdot} \dd \bm{\alpha} = \oint \bm{\nabla_{\alpha}} \phi \bm{\cdot} \dd \bm{\alpha} = \iint \bm{\nabla_\alpha} \times (\bm{\nabla_\alpha} \phi) \, \dd \alpha \, \dd \beta = 0 \, ,
\end{equation}
where we used Stokes theorem and the fact that the curl of a gradient always vanishes. Additionally, and importantly, if we constrain ourselves to spatially periodic flows, such as those represented by \eqref{expansions}, then we can choose a closed contour as in figure \ref{fig:Kelvin} which is a rectangle in label space with width $\lambda = 2\pi/k$, extending vertically from $(\beta=\beta_0)$ to the infinite bottom $(\beta \rightarrow -\infty)$ where the velocity vanishes. {\color{black}Because $\vec{A}$ is $\lambda$-periodic, which can be shown from \eqref{expansions}, the side contours cancel, and due to our deep water condition the bottom boundary does not contribute. Therefore the only segment in the contour which contributes is the top segment}, reducing \eqref{GammaLag} to
\begin{equation}\label{phaseavg}
    \Gamma = \int_{\alpha}^{\alpha + \lambda} \phi_\alpha \, \dd \alpha = \int_\alpha^{\alpha + \lambda} \dot{x} x_{\alpha'} + \dot{y} y_{\alpha'} \, \dd \alpha' = 0 \, ,
\end{equation}
or equivalently, that the phase average of $\dot{x}x_\alpha + \dot{y}y_\alpha$ is zero for any irrotational and horizontally periodic fluid. From here we can take advantage of our expansions \eqref{expansions} which relate $\alpha$ and $\tau$ derivatives as
\begin{equation}\label{derivativerelations}
    x_\alpha = \frac{c-\dot{x}}{c - U} \, , \qquad y_\alpha = -\frac{\dot{y}}{c - U} \, ,
\end{equation}
which transforms the integrand to
\begin{equation}
    \dot{x}x_\alpha + \dot{y}y_\alpha = \frac{c \dot{x} - (\dot{x}^2 + \dot{y}^2)}{c - U} \, .
\end{equation}
The consequence of \eqref{phaseavg} and \eqref{derivativerelations} is that
\begin{equation}
    (c-U)\Gamma = \int_{\alpha}^{\alpha + \lambda} c \dot{x} - (\dot{x}^2 + \dot{y}^2) \, \dd \alpha' = 0 \, ,
\end{equation}
which, defining a phase average with angle brackets $\langle \cdot \rangle$, yields
\begin{equation}
    c \langle \dot{x} \rangle = \langle \dot{x}^2 + \dot{y}^2 \rangle \, .
\end{equation}
A key realization is that in the Lagrangian frame
these quantities have physical interpretations such as $\langle \dot{x} \rangle$, the phase averaged momentum density, and $\langle \dot{x}^2 + \dot{y}^2 \rangle$, twice the phase averaged kinetic energy density, exactly how they would look in classical physics. Note that crucially, from expansions of the form \eqref{expansions}, the phase averaged momentum density is exactly the mean Lagrangian drift $U(\beta)$. Thus, we obtain the exact relation
\begin{equation}\label{mainresult}
    c U(\beta) = 2 \langle T \rangle \, ,
\end{equation}
where $T = \tfrac{1}{2}(\dot{x}^2 + \dot{y}^2)$ is defined as the kinetic energy density. This relation implies that the mean Lagrangian drift is linked to the kinetic energy of the system through the phase speed $c$, with all of its nonlinear corrections, implying that any irrotational wave of the form \eqref{expansions} that has energy must also have a mean Lagrangian flow, even if the underlying system does not represent surface gravity waves. Recall that all we have invoked here is Kelvin's circulation theorem, irrotational flow, and trajectories of the form \eqref{expansions}. 
Lastly, we emphasize that this relationship holds level-wise (i.e. for each vertical $\beta$ level) and as such encodes depth dependence. A similar relationship linking momentum density to kinetic energy density in the Eulerian frame was first found by \citet{Levi1924Book}, written in the form
\begin{equation}
    c I = 2 K \, ,
\end{equation}
where
\begin{equation}\label{impulse}
    I \equiv \overline{\int_{-h}^\eta u \, \dd y} \,
\end{equation}
is defined as the wave impulse, where $\eta$ is the sea surface, $-h$ the constant depth, and where the overline represents an Eulerian average in $x$ over one wavelength. The Eulerian kinetic energy density $K$ is defined as
\begin{equation}
    K \equiv \overline{\int_{-h}^\eta \tfrac{1}{2}(u^2 + v^2) \, \dd y} \, .
\end{equation}
This was later found to hold between any two material contours, or equivalently streamlines in a co-moving frame, by \cite{Starr47}. While this equation is similar in scope to \eqref{mainresult},that there is a direct connection between momentum and kinetic energy, these terms mean different things in different frames. 
{\color{black}To see this, consider \eqref{impulse}, the Eulerian averaged Eulerian momentum over one period. One can split this integral at the still water level $y=0$ yielding
\begin{equation}
    I = \overline{\int_{-h}^0 u \, \dd y} + \overline{\int_{0}^{\eta(x,t)} u \, \dd y}\, .
\end{equation}
Since the first integral's endpoints are constants, we can move the Eulerian phase average inside the integral. But since $\overline{u} = 0$ for irrotational surface gravity waves, the first integral is identically zero. Thus the only net Eulerian momentum density exists in the second integral between the still water level and the sea surface, or in other words, between the crests and troughs. Geometrically speaking, this occurs because at physical locations between the crests and troughs, a fixed $(x,y)$ point is outside of the fluid part of the time and thus does not experience a full period within the fluid.
}
Putting aside the difficulty of dealing with points partly outside of the fluid domain, these conserved Eulerian quantities are not physically connected to the mass flux of particles, and one would not be able to isolate the mean Lagrangian drift from such an approach.
The relationship found above \eqref{mainresult} holds for each material line of constant $\beta$, and as such shows equivalent vertical dependence in $U(\beta)$ and $\langle T \rangle$, but it is also more connected to the classical meanings of terms such as momentum and kinetic energy densities, which in the Lagrangian frame directly encodes the mean Lagrangian drift $U(\beta)$.

\subsection{Drift, Mean Water Level, and Mean Pressure}

The last connection we explore is between the wave-induced mean Lagrangian drift, the mean water level (MWL), and the mean fluid pressure. The mean water level, $y_0(\beta)$ in equation \eqref{expansions} at first appears to lack motivation -- it would seem that setting it to zero would be most natural. The reason this is not the case is due to the fact that the Lagrangian and Eulerian mean water levels are different, as the Lagrangian mean sums over particles, which are not equally spaced in physical space. Mathematically, the mean water levels in each frame are given as
\begin{align}
    \text{MWL}_{\text{Eul}} &= \frac{1}{\lambda}\int_0^\lambda \eta(x,t) \, \dd x \, , \\
    \text{MWL}_\text{Lag} &= \frac{1}{\lambda}\int_0^\lambda y(\alpha,0,t) \, \dd \alpha = y_0(\beta)
\end{align}
where $\eta(x,t)$ is the typical Eulerian sea surface elevation function, equivalent to $y(\alpha(x,t),0,t)$ assuming one inverts the mapping from $\alpha$ to $x$. The mean water level in the Eulerian frame, due to mass conservation, is the same as the still water level, so it is typically set to zero. If we try to convert this to the Lagrangian frame, we see that
\begin{equation}
    \text{MWL}_\text{Eul} = \frac{1}{\lambda} \int_0^\lambda \eta(x,t) \, \dd x = \frac{1}{\lambda} \int_0^\lambda y(\alpha,0,t) \frac{\partial x}{\partial \alpha} \dd \alpha  \neq \text{MWL}_\text{Lag} \, ,
\end{equation}
and thus the mean water levels are not the same. Precisely, they differ within the integral by the factor $x_\alpha$, which corresponds to the unequal spacing of particles along the water surface. In surface gravity waves this tends to bunch particles towards the wave crest and spread them out within the trough. A consequence of this is that the wave crests and troughs, as defined by Lagrangian phase in \eqref{expansions} are also of unequal lengths in physical space. The incompressibility condition in the Lagrangian frame is therefore maintained by stretching or compressing of particles in the vertical. We choose $y_0(\beta)$ such that the mean water level is zero, that is
\begin{equation}
    \langle y x_\alpha \rangle \big|_{\beta = 0} = 0 \, ,
\end{equation}
subject to the incompressibility condition \eqref{continuity} and and irrotational flow \eqref{"Vorticity"} which sets the vertical dependence. The physical meaning of the Lagrangian mean water level is that the presence of waves raises the average potential energy of the fluid parcels relative to their rest state in a still fluid. How is it then connected to the kinetic energy, and therefore the drift?

Following \cite{Pizzo2023}, by multiplying \eqref{"xEuler"} by $x_\beta$ and \eqref{"yEuler"} by $y_\beta$, we find
\begin{equation}
    p_\beta + g y_\beta + \ddot{y} y_\beta + \ddot{x}x_\beta = 0 \, .
\end{equation}
Taking advantage once again of our expansions \eqref{expansions}, we can write
\begin{equation}
    \ddot{y} = -(c-U)\dot{y}_\alpha \, , \quad \ddot{x} = -(c-U) \dot{x}_\alpha \, ,
\end{equation}
yielding
\begin{equation} \label{preVorticity}
    p_\beta + gy_\beta - (c-U)(\dot{y}_\alpha y_\beta + \dot{x}_\alpha x_\beta ) = 0 \, .
\end{equation}
Noting that the terms in parentheses are part of the vorticity, we can use our incompressibility condition \eqref{"Vorticity"} to write
\begin{equation}
    p_\beta + g y_\beta - (c-U)(\dot{x}_\beta x_\alpha + \dot{y}_\beta y_\alpha ) = 0 \, ,
\end{equation}
which, after another conversion between time and space derivatives, becomes
\begin{equation}
    \Big( p + g y + \frac{1}{2}(\dot{x} - c)^2 + \frac{1}{2}\dot{y}^2\Big)_\beta = 0 \, .
\end{equation}
Performing an indefinite integral of this equation yields

\begin{equation}\label{Bernoulli}
    p + g y + \frac{1}{2}(\dot{x}^2 + \dot{y}^2) - c \dot{x} + \frac{c^2}{2} = f(\alpha,\tau) \, ,
\end{equation}
where $f(\alpha,\tau)$ is a constant of integration, whose value is constrained by the boundary conditions. For our system, as we approach the infinite bottom, all wave terms $(\dot{x},\dot{y},U,y_0)$ vanish, and pressure becomes hydrostatic $p \rightarrow -g \beta$, which implies $f(\alpha,\tau) = c^2 / 2$. Clearly, this looks like Bernoulli's equation in Lagrangian coordinates as was first noticed by \cite{Pizzo2023}. If we take the phase average of this equation, we find
\begin{equation}
    \langle p \rangle + g \beta + g y_0(\beta) + \langle T \rangle - c U(\beta) = 0 \, ,
\end{equation}
an exact relation which holds level-wise. If we substitute the main result from the last section \eqref{mainresult}, this becomes
\begin{equation}
    \langle p \rangle + g \beta + g y_0(\beta)  + \langle T \rangle - 2 \langle T \rangle = 0 \, ,
\end{equation}
or rewritten,
\begin{equation} \label{Lagrangian}
    \langle p \rangle = \langle T \rangle - \langle V \rangle \, ,
\end{equation}
where $\langle V \rangle = g(\beta + y_0(\beta))$ is the average potential energy of particles. Thus the mean pressure acts as a Lagrangian for the system, which is similar to what \cite{luke1967} found in the Eulerian frame, {\color{black}where the expression of the pressure from Bernoulli's equation was used as a Lagrangian to get both the equations of motion and boundary conditions for the velocity potential $\phi$ and surface $\eta$}. If we apply Whitham's method using the averaged Lagrangian, and substitute the expansions \eqref{expansions} for $\langle T \rangle$ and $\langle V \rangle$, the action becomes
\begin{equation}
    \mathcal{A} = \int_{t_1}^{t_2}\int_{-\infty}^0 \frac{1}{2}U(\beta)^2 + \frac{1}{4}\sum_n n^2 k^2 (x_n + y_n)^2 (c-U(\beta))^2 - gy_0(\beta)  \, \dd \beta \, \dd \tau \, .
\end{equation}
Varying the mean Lagrangian drift itself yields
\begin{equation}
    \delta U : \qquad U(\beta) = \dfrac{\displaystyle \frac{c}{2} \sum  n^2 k^2(x_{n}^2 + y_{n}^2)}{\displaystyle 1 + \frac{1}{2} \sum n^2 k^2 (x_{n}^2 + y_{n}^2)} \, ,
\end{equation}
exactly the same as \eqref{"drift"} which was originally found by a dynamic constraint. 

One interesting consequence of 
\eqref{Lagrangian} is that it shows that the magnitude of the mean kinetic energy, mean potential energy, and mean pressure are all related. Recalling the last section, this equivalently means that the mean Lagrangian drift, the mean water level, and the mean pressure are also related. Substituting our forms for $\langle T \rangle $ and $\langle V \rangle$ gives
\begin{equation}\label{meanP}
    \langle p \rangle - (-g \beta) = \frac{c U(\beta)}{2} - g y_0(\beta) \, ,
\end{equation}
where $-g \beta$ is just the hydrostatic component of the pressure. At the surface $(\beta = 0)$, we know that the pressure vanishes via the dynamic boundary condition, which implies
\begin{equation}
    g y_0(0) = \frac{c U(0)}{2} \, ,
\end{equation}
a result first discovered by \cite{LH1986}, though only at the surface. Our equation \eqref{meanP} implies that at each and every material line, the balance between mean kinetic energy (drift) and mean potential energy (MWL) differs exactly by the mean pressure deviation at that depth, which is in general nonzero. This result connects the mean water level, a purely geometric quantity, to the mean momentum and pressure, dynamic quantities, and as such we show how one can infer dynamics from geometry, and vice versa, for irrotational water waves. 

\section{Conservation Laws}

The previous section introduced a close connection between momentum and energy for spatially periodic, irrotational waves in a fluid. What are the corresponding conservation laws for these quantities? Returning to the momentum equations
\begin{equation}\label{conservationX}
        \mathcal{J} \ddot{x} + p_\alpha y_\beta - p_\beta y_\alpha = 0 \, ,
\end{equation}
\begin{equation}\label{conservationY}
        \mathcal{J} \ddot{y} + p_\beta x_\alpha - p_\alpha x_\beta + \mathcal{J}g = 0  \, ,
\end{equation}
we can derive a conservation law for total horizontal momentum by vertically integrating \eqref{conservationX} from the infinite bottom to the free surface,
\begin{equation}
    \int_{-\infty}^0 \mathcal{J} \ddot{x} \, \dd \beta + \int_{-\infty}^0 p_\alpha y_\beta - p_\beta y_\alpha \, \dd \beta = 0 \, .
\end{equation}
Recognizing that incompressibility requires $\mathcal{J}$ to be time independent, we can pull a time derivative out of the first integral. In addition, if we consider the integral
\begin{equation}
    \frac{\d}{\d \alpha}\int_{-\infty}^0 p y_\beta \, \dd \beta = \int_{-\infty}^0 p_\alpha y_\beta \, \dd \beta + \int_{-\infty}^0 p y_{\alpha \beta} \, \dd \beta \, ,
\end{equation}
and apply integration by parts on the last term, we obtain the expression
\begin{equation}\label{xmomconserved}
    \frac{\d}{\d \tau} \underbrace{\int_{-\infty}^0 \mathcal{J}\dot{x} \, \dd \beta }_{\equiv \mathcal{I}}+ \frac{\d}{\d \alpha}\underbrace{\int_{-\infty}^0 p y_\beta \, \dd \beta}_{\equiv S} = p y_\alpha \Big|_{b = 0} \, ,
\end{equation}
where we define $\mathcal{I}$ and $S$ as the vertically integrated horizontal momentum density and flux, respectively.
Put in this way, \eqref{xmomconserved} becomes a standard conservation law for bulk horizontal momentum,
\begin{equation}\label{xconservation}
    \frac{\d \mathcal{I}}{\d \tau} + \frac{\d S}{\d \alpha} = p y_\alpha \Big|_{\beta = 0} \, ,
\end{equation}
where $py_\alpha$ at the surface is the source of momentum.
Note that
\begin{equation}
    p y_\alpha \Big|_{\beta = 0} = p \eta_x x_\alpha = p \eta_x (1 + \ldots)\Big|_{\beta = 0} \, ,
\end{equation}
since $\eta(x,t) \equiv y(\alpha(x,t),0,t)$.
In the Eulerian frame, the source of momentum from the wind is given to lowest order by the correlation of surface pressure and sea surface slope $\eta_x$, which we see validated here \citep{Miles1957, Phillips1977}.
Recall that for an unforced wave, $p=0$ at the surface, and total momentum is conserved. 

We perform the same process for the vertically integrated energy by multiplying \eqref{conservationX} by $\dot{x}$ and \eqref{conservationY} by $\dot{y}$, adding the two equations and vertically integrating to get
\begin{equation}\label{Econservation}
    \frac{\d E}{\d \tau} + \frac{\d F}{\d \alpha} = p(\dot{x}y_\alpha - x_\alpha \dot{y}) \Big|_{\beta = 0} \, ,
\end{equation}
where $E$ is defined as the vertically integrated energy density
\begin{equation}
    E \equiv \int_{-\infty}^0 \mathcal{J}\bigg( \frac{\dot{x}^2 + \dot{y}^2}{2} + g y \bigg) \, \dd \beta \, ,
\end{equation}
and $F$ is defined as the vertically integrated energy flux
\begin{equation}
    F \equiv \int_{-\infty}^0 p(x_\beta \dot{y} - \dot{x}y_\beta) \, \dd \beta \, .
\end{equation}
Just as with horizontal momentum, if pressure vanishes at the surface, then total energy is conserved.

In the Lagrangian frame, the average momentum density $\langle \mathcal{I} \rangle$ is all contained within the mean Lagrangian drift, as it is the only term that survives the phase averaging.
If there was no pressure forcing and we phase-averaged the horizontal momentum conservation law \eqref{xconservation}, we would find
\begin{equation}
    \frac{\d \langle \mathcal{I} \rangle}{\d \tau} = 0 \, ,
\end{equation}
which just states that the total integrated Lagrangian mean drift, or equivalently the average horizontal momentum density, is conserved.
To see how this momentum (and therefore energy) can increase in time, we need to allow for a nonzero pressure forcing, which leads naturally to an example of generating Stokes waves from rest.

\subsection{Generating Stokes waves from rest}

While the previous analysis showed why a drift must occur if the wave is progressive, irrotational and contains energy, it is helpful to also show how a mean Lagrangian drift can arise on an initially quiescent flow.
To begin, we consider a still fluid which at $\tau=0$ is subject to an external wavelike surface pressure forcing (e.g. by wind)
\begin{equation}
    p(\beta = 0) = \epsilon p_0 \sin(k \alpha - \omega \tau) \, ,
\end{equation}
where $\epsilon \ll 1$ is our small parameter, and we take $\omega = \sqrt{g k}$ so that the pressure disturbance propagates at the same speed as a surface gravity wave with the same wavelength.
Physically speaking, this pressure forcing drives a resonant response in the sea surface, generating waves whose amplitudes grow linearly in time {\color{black}\citep{pizzowagner2021}}. In the Lagrangian frame, the particle trajectories for this system valid to second order in $\epsilon$ are found to be
\begin{align}
    x(\alpha,\beta,\tau) &= \alpha + \frac{\epsilon \omega p_0}{2 g}\tau e^{k \beta}\sin(k \alpha - \omega \tau) + \frac{\epsilon^2 p_0^2 k^2 \omega}{12 g}e^{2k\beta}\tau^3 \, \label{xforced}, \\
    y(\alpha,\beta,\tau) &= \beta - \frac{\epsilon \omega p_0}{2 g}\tau e^{k \beta}\cos(k \alpha - \omega \tau)+ \frac{\epsilon^2 p_0^2 k^2}{8 g}e^{2k\beta}\tau^2 \, , \label{yforced} \\
    p(\alpha,\beta,\tau) &= -g \beta + + \epsilon p_0 e^{k \beta}\sin(k \alpha - \omega \tau) - \frac{\epsilon^2 p_0^2 k}{8 g}(e^{2k\beta} - 1) \label{pforced}\, ,
\end{align}
{\color{black} where solutions are found through a standard perturbation approach {\color{black}\citep[see][ch 1. for an outline of the method]{Salmon2020}}}. Note that we have ignored the mean Lagrangian drift in the phase since it does not affect the results to second order. Because the amplitude grows linearly in time, the mean Lagrangian drift which is normally proportional to the square of the amplitudes times $\tau$ correspondingly grows as $\tau^3$.
The second order term in \eqref{yforced} is just the mean water level which scales as the square of the amplitude, which is there to ensure the incompressibility condition subject to the gauge $\mathcal{J} = 1$. To see how this explicitly connects to the drift, we first will perform a phase average of the horizontal momentum conservation law \eqref{xconservation}
%
\begin{equation}
    \frac{\d}{\d \tau} \int_{-\infty}^0 \langle \dot{x} \rangle \, \dd \beta = \frac{\d}{\d \tau} \int_{-\infty}^0  U(\beta,\tau) \, \dd \beta = \langle p y_\alpha \big|_{\beta = 0} \rangle \, . \label{avgmom}
\end{equation}
This gets rid of the flux terms since the entire solution is spatially periodic in $\alpha$.
Our gauge choice $\mathcal{J} = 1$ trivializes the Jacobian term. 
Thus the mean external pressure forcing, represented by the correlation of $p$ and $y_\alpha$ at the surface, provides a source of horizontal momentum which fuels the increase of the mean horizontal momentum, or equivalently the vertically integrated Lagrangian mean drift.

Inserting our solutions \eqref{xforced} -- \eqref{pforced} into the phase averaged  horizontal momentum conservation law \eqref{avgmom}, we confirm our results
\begin{equation}
    \frac{\d}{\d \tau} \int_{-\infty}^0 U(\beta,\tau) \, \dd \beta = \langle p y_\alpha \big|_{\beta = 0} \rangle = \frac{\epsilon^2 p_0^2 \omega k }{4 g} \tau  \, ,
\end{equation}
notably that the mean momentum input from the wind to the waves in order to generate wave growth goes entirely into increasing the mean Lagrangian drift.

Though the simple example presented above is by no means intended to be a complete description of how waves are generated, it illustrates a physical source for the mean Lagrangian flow.
There need not be any small-amplitude approximations either; equation \eqref{xconservation} only assumes inviscid flow and infinite depth.

In short, to generate periodic irrotational surface gravity waves from rest, there must be a mean input of horizontal momentum to the water, or equivalently a convergence of momentum flux.
This mean momentum lives entirely within the mean Lagrangian flow, identifying the wave induced mean Lagrangian drift as nothing more than the average horizontal momentum necessary for the generation of irrotational surface gravity waves by any conservative force.

As a further check, we see by inserting our solutions \eqref{xforced} -- \eqref{pforced} into the phase average of the energy conservation law \eqref{Econservation}, we recover

\begin{equation}
    \frac{\d \langle E\rangle}{\d \tau} = \Big\langle p(\dot{x}y_\alpha - x_\alpha \dot{y})\Big|_{\beta = 0}\Big \rangle  = \frac{1}{4}\epsilon^2 k p_0^2 \tau + O(\epsilon^4) \, ,
\end{equation}
which shows that, just as with momentum, generating waves requires a flux of energy from the wind to the waves.
However to lowest order, this energy resides wholly in the orbital particle motion and gravitational potential energy from the mean water level.
Recalling the result from the previous section, we do indeed see that the kinetic energy and mean Lagrangian drift are related, notably that the mean source of momentum multiplied by $c =\sqrt{g/k}$ is equivalent to the mean source of energy at this order, once again highlighting the connection between these two quantities.

\section{Discussion}

In this paper we showed that the mean Lagrangian drift, equivalent to the phase averaged momentum density in the physically-motivated Lagrangian frame, is intimately connected to the vorticity and energy densities for irrotational, monochromatic and spatially periodic waves. We further highlighed this connection by showing that sources of momentum and energy (e.g. from the wind) all add to the momentum and energy of the wave field using a simple example of an initially quiescent fluid resonantly forced by a wavelike pressure disturbance at the surface. Physically speaking, this implies that the wave induced mean Lagrangian drift arises due to the necessary input momentum and energy to generate an irrotational wave from rest by any conservative force. {\color{black} Thus we are well equipped to answer the questions posed in the introduction. Permanent, progressive and irrotational waves require a mean motion of water due to the fact that for these waves to have kinetic energy, they require a net momentum (or mass flux), which in the Lagrangian frame resides in the mean Lagrangian drift. Its magnitude and direction are set by the strict dynamic constraint of irrotational flow, as prescribing the vorticity on particles is equivalent to prescribing their mean Lagrangian drift (see the appendix for cases with nonvanishing vorticity). Finally, the mean Lagrangian drift is not simply related to the mean kinetic energy density in a bulk sense, it is exactly proportional to it, at all vertical material levels, with a factor of $1/c$.}

Our theoretical results imply that for irrotational, monochromatic and periodic waves, the mean kinetic energy and momentum of particles are intimately related through the wave's phase speed. This could suggest that anywhere energy is jettisoned, such as by wave breaking, it is accompanied by a shedding of momentum {\color{black}density} to the underlying mean flow \citep{Rapp1990}. This connection between wave energy and mean momentum helps to illuminate the close two-way coupling between currents and waves, and as such will be of particular interest to the air-sea interaction community seeking to model transport and energy budgets between the atmosphere and ocean. 

In addition, we explored a connection between the mean potential energy, mean kinetic energy, and mean pressure, showing that at the surface, mean kinetic and mean potential energies are equal, which relates the surface drift directly to the mean water level. This is especially relevant to the observational community as direct measurements of the mean Lagrangian drift are particularly difficult \citep{kenyon1969}, especially close to the surface \citep{lenain2020}.  On the other hand, measurements of geometric properties such the Lagrangian mean water level might offer an alternative way to estimate the mean Lagrangian drift{\color{black}, as in \cite{McAllisterVanDenBremer2019}}. {\color{black} One added caveat is that all of this analysis rests upon the assumption of two-dimensional flow, which is fundamentally different than fully three-dimensional flow, due to the fact that vorticity is no longer conserved on fluid particles because of vortex tilting/stretching. Examples of how to employ the Lagrangian frame for three-dimensional flow can be found in \cite{YAKUBOVICH_ZENKOVICH_2001}.}

It should be noted that the results presented here are similar to the ``pseudomomentum rule" in Generalized Lagrangian Mean (GLM) theory, which states that $O(A^2)$ mean forces can be calculated as if pseudomomentum were momentum, and the fluid medium were absent \citep{mcintyre2019}, where $A$ is the small wave amplitude. For surface waves, pseudomomentum per unit mass is defined as the wave energy over the phase speed $c$. We emphasize that it is twice the mean \emph{kinetic} energy, which when divided by $c$, yields $U(\beta)$, valid to all orders of amplitude. To lowest order, the mean kinetic and potential energies are equal, which explains the $O(A^2)$ result. The exact difference between mean kinetic and potential energy is given by the mean pressure \eqref{Lagrangian}, which has nonvanishing terms starting at $O(A^4)$. {\color{black} The field of wave-mean interactions is vast \citep{Buhler2014,Leibovich1983,Thomas_2016}, and while we do not investigate a general connection between momentum, energy, and vorticity for different types of waves, {\color{black}one may look towards \cite{Salmon_2016,Wagner_Young_2015,Thomas2023} for other examples}, a purely Lagrangian framework may prove insightful to such systems}.

There is also a connection between these results for irrotational waves and the Darwin drift for irrotational flow around a submerged body \citep{Darwin1953}. Darwin's result states that the `added mass' of a body moving through an irrotational fluid, which is related to the kinetic energy of the body, is equal to the `drift volume' swept out by the passing of the object. The equivalent drift volume for surface waves is simply a vertical integral of the Lagrangian mean drift, which by \eqref{mainresult} is directly related to the kinetic energy of the waves. The similaritites between Darwin drift and Stokes drift were first explored by \cite{eamesMcintyre1999}.

Note that these results do not hold in the general case of rotational waves, such as \cite{Gerstner1802} waves, which have no mean Lagrangian drift and therefore no net Lagrangian momentum density, yet still have a non-zero energy from their orbital motion. In this case, the connection between kinetic energy and drift fails due to the non-vanishing circulation. Here we focused solely on irrotational flow, though if desired any arbitrary vorticity could be prescribed to the system, generating a nonzero circulation which would balance the kinetic energy term in lieu of the drift. A comprehensive investigation of rotational flow is explored in the appendix.

Lastly, this work highlights the benefits of working directly within the Lagrangian frame, which is the most natural way to compute and interpret fundamentally Lagrangian quantities. Some of them, such as momentum and energy, take on more classical meanings when computed in this frame, and as such can be easier to interpret.

\small
\vspace{3mm}
\noindent \textbf{Acknowledgements.} 
{We thank the three anonymous referees whose constructive comments significantly improved the manuscript. We thank R. Salmon for helpful comments.}\\

\noindent \textbf{Funding.}
{Blaser, Pizzo, and Lenain were partially supported by NSF OCE-2219752 and 2342714, NASA 80NSSC19K1037 (S-MODE) and 80NSSC23K0985 (OVWST) awards. Villas Bôas was supported by NASA 80NSSC23K0979 (OVWST) and NASA 80NSSC24K0411 (S-MODE) awards.}\\

\noindent \textbf{Declaration of interests.}
{The authors report no conflict of interest.}
\vspace{3mm}\\
\noindent \textbf{Author Contributions.}\\
\textbf{Blaser} Formal analysis (lead); writing - original draft presentation (lead); writing - review and editing (lead); conceptualization (supporting). \textbf{Benamran} Formal analysis (supporting); writing - original draft presentation (supporting); conceptualization (supporting). \textbf{Villas Bôas} Conceptualization (supporting); writing - review and editing (supporting). \textbf{Lenain} Conceptualization (supporting); Funding Acquisition (lead for AB and RB); Supervision (supporting);  Writing - review and editing (supporting); \textbf{Pizzo} Conceptualization (lead); Formal analysis (supporting); Writing - review and editing (supporting); Funding Acquisition (supporting); Supervision (lead). 
\\

\noindent \textbf{Author ORCIDs.}\\
\orcidlink{0000-0002-3104-4589} Aidan Blaser \href{https://orcid.org/0000-0002-3104-4589}{https://orcid.org/0000-0002-3104-4589} \\
\orcidlink{0009-0000-2205-3443} Rapha{\"e}l Benamran \href{https://orcid.org/0009-0000-2205-3443}{https://orcid.org/0009-0000-2205-3443} \\
\orcidlink{0000-0001-6767-6556} Ana B. Villas Bôas \href{https://orcid.org/0000-0001-6767-6556}{https://orcid.org/0000-0001-6767-6556} \\
\orcidlink{0000-0001-9808-1563} Luc Lenain \href{https://orcid.org/0000-0001-9808-1563}{https://orcid.org/0000-0001-9808-1563}\\
\orcidlink{0000-0001-9570-4200} Nick Pizzo \href{https://orcid.org/0000-0001-9570-4200}{https://orcid.org/0000-0001-9570-4200}

\normalsize
\appendix
\section{Rotational Waves}

{\color{black} Due to the relative simplicity of irrotational planar flow in the Eulerian frame, allowing for the use of a velocity potential, the vast majority of surface wave literature concerns irrotational waves. However it is generally understood that common effects such as wave breaking \citep{Rapp1990,Pizzo2013} and ocean-atmosphere shear flows which can generate surface waves \citep{Young2014} imply that real world surface gravity waves often contain vorticity. Without the benefit of potential flow, Eulerian treatments of rotational surface gravity waves \citep{PHILLIPS_2001,CONSTANTIN_SATTINGER_STRAUSS_2006} become quite difficult. This is where the Lagrangian machinery outlined above operates best as vorticity is conserved on particles in two-dimensional inviscid flow. }Thus any Lagrangian formulation presented above which holds for a various collection of particles can be easily modified to account for vortical waves. As an example, we direct the reader to \cite[][eq. 2.9]{Pizzo2023} to see how the drift is dynamically constrained by an arbitrary vorticity. For convenience we rewrite it here
\begin{equation}\label{nickdrift}
    U(\beta) = \dfrac{\displaystyle \frac{c}{2}\sum_n n^2 k^2 (x_n^2 + y_n^2) - \int_{-\infty}^\beta \langle \mathcal{J}q \rangle \, \dd \beta'}{\displaystyle 1 + \frac{1}{2}\sum_n n^2 k^2 (x_n^2 + y_n^2)} \, .
\end{equation}

To determine the balance between mean momentum, kinetic energy, and vorticity for water waves, we return to the definition of the circulation in the Lagrangian frame \eqref{gammaDef}, and apply Stokes' theorem to find
\begin{equation}
    \Gamma = \oint \bm{A} \cdot \dd \bm{\alpha}  = \iint (\bm{\nabla}_{\bm{\alpha}} \times \bm{A}) \cdot \bm{\hat{n}} \, \dd \alpha \,  \dd \beta \, ,
\end{equation}
where $\bm{\hat{n}}$ is the unit normal. For the contour used above described by figure \ref{fig:Kelvin}, due to the clockwise orientation, $\bm{\hat{n}}$ points into the page and we have
\begin{equation}
    \Gamma = - \int_{-\infty}^0 \int_{\alpha}^{\alpha + \lambda} \mathcal{J} q \, \dd \alpha' \, \dd \beta' \, ,
\end{equation}
using the definition of $\bm{A}$ and $q$. {\color{black} This is also equivalent to the enclosed vorticity (with a minus sign for the orientation of $\bm{\hat{n}}$) within the material loop, identical to that in the Eulerian frame.} Since the $\alpha$ derivative is over a wavelength, we can convert this to a phase average, resulting in
\begin{equation}\label{circApp}
    \Gamma = \oint \bm{A} \cdot \dd \bm{\alpha} = \int_\alpha^{\alpha + \lambda} \dot{x}x_\alpha + \dot{y}y_\alpha \, \dd \alpha' = - \int_{-\infty}^\beta \langle \mathcal{J} q \rangle \, \dd \beta' \, ,
\end{equation}
for our chosen material loop, since as before, the side and bottom contours do not contribute. The contour integral part of the equation is unchanged from the irrotational case, so after a few manipulations, the result becomes
\begin{equation}\label{vorticityResult}
    c U(\beta) - 2 \langle T \rangle = -(c-U(\beta)) \int_{-\infty}^\beta \langle \mathcal{J}q \rangle \, \dd \beta' \, ,
\end{equation}
where $U(\beta)$ and $\langle T \rangle$ are defined the same as before. As a quick check consider the \cite{Gerstner1802} wave, which is exactly described by the circular trajectories and pressure
\begin{align}
    x(\alpha,\beta,\tau) &= \alpha - A e^{k \beta}\sin(k (\alpha - c \tau)) \, , \\
    y(\alpha,\beta,\tau) &= \beta + A e^{k \beta}\cos(k(\alpha - c \tau)) + \frac{1}{2}A^2 k \, , \\
    p(\beta,\tau) &= -g\beta + \frac{1}{2}A^2 k^2 c^2 (e^{2 k \beta} - 1) \, ,\label{Gerstner}
\end{align}
where $A<1$ is the amplitude of the wave, and $c = \sqrt{g/k}$ is the exact phase speed. These waves have vorticity, but no mean Lagrangian drift $(U = 0)$. Inserting these into \eqref{vorticityResult} yields
\begin{equation}
    2 \langle T \rangle = c \int_{\infty}^\beta \langle \mathcal{J} q \rangle \, \dd \beta' = A^2 k^2 c^2 e^{2 k \beta}\, ,
\end{equation}
which validates the result for this special case. Thus \eqref{vorticityResult} is the {\color{black}generalization} of \eqref{mainresult}, and states that there is actually a balance between drift, kinetic energy density, and vorticity for the waves considered. Stokes waves, where the balance is entirely between drift and kinetic energy density, or Gerstner waves, which balance kinetic energy density and vorticity, are thus limiting cases for this general result.


Finally, we investigate how the Bernoulli equation \eqref{Bernoulli} is altered by allowing for an arbitrary vorticity starting at equation \eqref{preVorticity} to find
\begin{equation}
    p_\beta + g y_\beta - (c-U)(\mathcal{J} q + \dot{x}_\beta x_\alpha + \dot{y}_\beta y_\alpha ) = 0 \, ,
\end{equation}
\begin{equation}
    \Big( p + g y + \frac{1}{2}(\dot{x}-c)^2 + \frac{1}{2}\dot{y}^2\Big)_\beta - (c-U)\mathcal{J}q = 0 \, .
\end{equation}
Once again, we can integrate this equation and use the same argument to constrain the constant of integration $(f(\alpha,\tau) = c^2/2)$ to find
\begin{equation}
    p + gy + \frac{\dot{x}^2 + \dot{y}^2}{2} - c \dot{x} = \int_{-\infty}^\beta (c-U)\mathcal{J}q \, \dd \beta' \, ,
\end{equation}
a result also derived in \cite{Pizzo2023}. If we now phase average this equation, we get
\begin{equation}
    \langle p \rangle + g \beta + g y_0(\beta) + \langle T \rangle - c U = \int_{-\infty}^\beta (c-U)\langle \mathcal{J} q \rangle \, \dd \beta \, .
\end{equation}
Using our new result linking drift, kinetic energy density and vorticity \eqref{vorticityResult}, we can write the mean pressure as
\begin{align}
    \langle p \rangle &= \langle T \rangle - \langle V \rangle  - (c-U)\int_{-\infty}^\beta \langle \mathcal{J}q \rangle \, \dd \beta' + \int_{-\infty}^\beta (c-U)\langle \mathcal{J}q \rangle \, \dd \beta' \, , \\
    & = \langle T \rangle - \langle V \rangle - \int_{-\infty}^\beta \frac{\d U(\beta')}{\d \beta}\Gamma(\beta') \, \dd \beta' \, , 
\end{align}
where we use the definition of the circulation $\Gamma$ as above \eqref{circApp}. This result shows how, when vorticity is present, the mean pressure is not precisely equal to $\langle T \rangle - \langle V \rangle$, and differs by terms related to the vorticity and the mean Lagrangian drift. If we again insert this into an averaged Lagrangian via Whitham's method and vary $U(\beta)$, we recover \eqref{nickdrift}. Interestingly, when either the vorticity is zero, as in \eqref{Lagrangian}, or when the mean Lagrangian drift is zero, as in a Gerstner wave, we do in fact see that the mean pressure acts as a Lagrangian for the system, i.e.
\begin{equation}
    \langle p \rangle_{\text{Stokes}} = \langle T \rangle - \langle V \rangle \, , \quad \langle p \rangle_{\text{Gerstner}} = \langle T \rangle - \langle V \rangle \, ,
\end{equation}
but not necessarily for intermediate waves with nonzero drift and vorticity. Writing this result explicitly in terms of the mean Lagrangian drift and mean water level results in
\begin{equation}
    \langle p \rangle - \langle -g \beta \rangle = \frac{c U(\beta)}{2} - gy_0(\beta) - \frac{c-U}{2}\Gamma(\beta) - \int_{-\infty}^\beta \frac{\partial U(\beta')}{\d \beta} \Gamma(\beta') \, \dd \beta' \, .
\end{equation}
For the Gerstner wave, where $U$ vanishes, we can use \eqref{Gerstner} to write
\begin{equation}
    \frac{1}{2}A^2 k^2 c^2 (e^{2 k \beta} - 1) = -g y_0(\beta) - \frac{c}{2}\Gamma(\beta) \, ,
\end{equation}
which, by computing the circulation, yields
\begin{equation}
    y_0 = \frac{1}{2}A^2 k \, ,
\end{equation}
independent of depth. This is in contrast to the irrotational case, whose mean water level decays exponentially, highlighting again the importance of vorticity for these quantities. At the surface, where the mean pressure vanishes for an unforced wave, we have
\begin{equation}
    g y_0(0) = \frac{c U(0)}{2} - \frac{c-U(0)}{2}\Gamma(0) - \int_{-\infty}^0 \frac{\d U(\beta')}{\d \beta}\Gamma(\beta') \, \dd \beta'
\end{equation}
and as such, the mean water level, a purely geometric quantity, is related to both the dynamic mean Lagrangian drift and the vorticity.

\bibliographystyle{jfm}
\bibliography{ref}

\end{document}